\documentclass[preprint]{aastex}

\begin{document}

\title{Dust properties in afterglow of GRB 071025 at z$\thicksim$5}

\author{Minsung Jang\altaffilmark{1}, Myungshin Im\altaffilmark{1}, Induk Lee\altaffilmark{2}, Yuji Urata\altaffilmark{2}, Lijin Huang\altaffilmark{3}, Hiroyuki Hirashita\altaffilmark{3}, Xiaohui Fan\altaffilmark{4} and Linhua Jiang\altaffilmark{4}}

\altaffiltext{1}{Center for the Exploration of the Origin of the Universe (CEOU), Astronomy Program, Department of Physics \& Astronomy, Seoul National University, Shillim-Dong, Kwanak-Gu, Seoul 151-742, Republic of Korea; msjang.astro@gmail.com, mim@astro.snu.ac.kr}
\altaffiltext{2}{Institute of Astronomy, National Central University, Chung-Li 32054, Taiwan}
\altaffiltext{3}{Institute of Astronomy and Astrophysics, Academia Sinica, PO Box 23-141, Taipei 10617, Taiwan}
\altaffiltext{4}{Steward Observatory, University of Arizona, USA}

\begin{abstract}
At high redshift, the universe was so young that core-collapse supernovae (SNe) are suspected to be the dominant source of dust production. However, some observations indicate that the dust production by SNe is an inefficient process, casting doubts on the existence of abundant SNe-dust in the early universe. Recently, \citet{per10} reported that the afterglow of GRB 071025 -- an unusually red GRB at $z \sim 5$ -- shows evidence for the SNe-produced dust. Since this is perhaps the only high redshift GRB exhibiting compelling evidence for SNe-dust but the result could easily be affected by small systematics in photometry, we re-examined the extinction properties of GRB 071025 using our own optical/near-infrared data at a different epoch. In addition, we tested SNe-dust models with different progenitor masses and dust destruction efficiencies to constrain the dust formation mechanisms. By searching for the best-fit model of the afterglow spectral energy distribution, we confirm the previous claim that the dust in GRB 071025 is most likely to originate from SNe. We also find that the SNe-dust model of 13 or 25M$_{\odot}$ without dust destruction fits the extinction property of GRB 071025 best, while pair-instability SNe (PISNe) models with a 170M$_{\odot}$ progenitor poorly fit the data. Our results indicate that, at least in some systems at high redshift, SNe with intermediate masses within 10 $-$ 30M$_{\odot}$ were the main contributors for the dust enrichment, and the dust destruction effect due to reverse shock was negligible.
\end{abstract}

\keywords{gamma ray : burst --- dust, extinction --- galaxies: high-redshift --- galaxies: ISM}

\section{Introduction}
Dust plays an important role in understanding the galaxy formation and evolution over the cosmic history. Dust absorbs and scatters ultraviolet (UV) and optical lights from young stars, making it difficult to comprehend how stars were formed in galaxies using the optical and the near-infrared (NIR) observations that sample the redshifted UV/optical lights of high redshift objects. Although the infrared and sub$mm$ observations allow us to get a glimpse of the light from young stars re-emitted in the rest-frame infrared (e.g. Choi et al. 2006), such observations for the high redshift universe are still technically challenging. Therefore, it is imperative to know how the dust extinguishes the UV/optical light at high redshift, in order to understand how stars and galaxies were formed in the early universe.

In the local and the intermediate redshift ($z < 3$) universe, Asymptotic Giant Branch (AGB) stars are one of the major dust sources \citep{fer06}.
 When the universe was much younger ($\thicksim$ 1 Gyr) or at $z > 5$, however, AGB stars cannot be the main producers of dust simply because stars at that epoch do not have enough time to evolve into the AGB-phase.

 It is also known that core-collapse supernovae (SNe) can produce dust grains
\citep{koz89, tod01,noz03}. Since core-collapse SNe can occur on short time scales, 
 the dust in the early universe is speculated to be formed from SNe \citep{mai04}.
 However, the dust production from core collapse SNe is known to be inefficient \citep{gre04,erc07,rho09}, making it unclear if core-collapse SNe can produce a sufficient amount of dust to significantly obscure star formation activities at high redshift.

Observations of broad absorption line quasars at high redshift and some gamma ray bursts (GRBs) suggest that their rest-frame spectral energy distributions (SEDs) can be fitted best if one assumes a dust extinction curve produced by SNe-dust (Maiolino et al. 2004; Hirashita et al. 2005, hereafter H05; Stratta et al. 2007; Gallerani et al. 2010). In such a study, GRBs can be greatly useful tools. A GRB afterglow spectrum in UV/optical follows a simple power law that results from the synchrotron radiation of accelerated relativistic electrons by internal and external shocks propagating into the surrounding medium of GRB progenitors, making it easy to trace dust extinction curves imprinted on the SEDs . The extreme brightness of GRBs helps us obtain the SEDs of them at high redshift even with a small telescope.

Currently, observational constraints on dust in high redshift GRBs are weak. While \citet{str07} claimed that the extinction property of GRB 050904 at $z = 6.29$ is due to SNe-dust, a re-analysis of the GRB 050904 data by \citet{zar10} reveals no strong evidence to support the SNe-dust. Other high redshift GRBs are also known to harbor little amount of dust \citep{cuc11, zar11}.

GRB 071025 is an unusually red GRB believed to be at a redshift of $z \sim 5$ (Im et al. 2007; Perley et al. 2010, hereafter P10). Its red color and the inflexed shape of the afterglow SED suggest dust extinction dominated by SNe-dust \citep{per10}. Other than GRB 050904, GRB 071025 is currently the only GRB which is claimed to harbor SNe-dust. However, as we saw from the case of GRB 050904, subtle systematic errors in photometry could lead to a large difference in the derived dust extinction properties (Zafar et al. 2010; P10). Therefore, in order to provide an independent assessment of the dust extinction properties of GRB 071025, we analyzed the SED of its afterglow using our own dataset of optical/NIR images taken in $RIJHK_{s}$ and constructed at a different epoch from P10. In addition to the verification for SNe-dust, we also examined a variety of SNe-dust extinction models from H05 and \citet{hir08} (hereafter H08) to place constraints on the progenitor mass and the importance of the destruction of the dust by external medium.

\section{Analysis}

\subsection{Observation and Data Analysis} \label{bozomath}

We started the follow-up imaging observation of the afterglow of GRB 071025 using the 1-m telescope at Mt. Lemmon Optical Astronomical Observatory (LOAO) on 2007 October 25, 04:26:54.07 UT, $\sim$1080 sec after the $Swift$ BAT trigger \citep{pag07}. We obtained $B$, $V$, $R$ and $I$ band imaging data till the end of October 25, and a part of October 26 \citep{im07, lee10}. Our data showed detections of a bright afterglow in the $I$ band and a faint counterpart in the $R$ band. The afterglow was not detected in the $B$ and $V$ bands. We reduced the data using the standard reduction procedures of dark subtraction and flat-fielding using IRAF packages. We obtained the photometry of the afterglow and surrounding stars using SExtractor with the parameters of 5 pixels for DETECT MINAREA, 1.2 $\sigma$ for DETECT THRESH, and 1.2 $\sigma$ for ANALYSIS THERSH \citep{ber96}. 
 The magnitudes are computed using circular apertures with an aperture radius of 1.5 FWHM of point spread functions
 and aperture corrections derived from bright stars around the afterglow which amount to 0.2 -- 0.4 mag.
We obtained the data for standard stars (SA95 and SA115) in the second night which were used to do the photometry calibration of the GRB 071025 field.

The $J$, $H$, and $K_{s}$ band data were obtained by the Simultaneous Quad Infrared Imaging Device (SQIID) on the Kitt Peak Mayall 4-meter telescope at Steward Observatory, with the start time of 2007 October 25, 05:56:59 UT \citep{jia07}. The total exposure time in each band was 240 second, and the three bands data were taken at the same time. The photometry was calibrated using 2MASS point sources brighter than $J$ = 16.0, $H$ = 15.5, and $K_{s}$ = 14.0 mags within 120 arcsec distance from the afterglow. Additional $J, H$, and $K_{s}$ data were obtained at $\Delta t \simeq 1.05$ days with the Canada-France-Hawaii Telescope.

 The results of the photometry are presented in Table 1, where the magnitudes are given in the AB system.
For the conversion of the Vega-based magnitudes into AB magnitudes, we used the conversion relation of \citet{fre94} for $RI$ bands, and \citet{cil05} for $JHK_{s}$ bands. Galactic extinction is also corrected using $E(B-V)$ = 0.07 \citep{sch98}, which amounts to 0.19, 0.14, 0.06, 0.04, 0.03 mags in $RIJHK_{s}$, respectively.

\subsection{Synchronization of Photometry} \label{bozomath}

The afterglow data were taken at different epochs but GRB afterglows change their brightness and in some cases their colors, making it necessary to construct a SED at a given epoch for the subsequent analysis of the afterglow. 
For this purpose, we use the mid-time of the $JHK_{s}$ data (6605 secs after the burst trigger) as the epoch for which we construct the SED, since the $JHK_{s}$ data were taken simultaneously. Note that we use external shock model with the form of  $F_{t,\nu} \sim t^{-\alpha}\,\nu^{-\beta}$ \citep{sar98} for the GRB 071025 afterglow, where $t$ is the delay time since the BAT trigger, $\nu$ is the obseved frequency, $\alpha$ is temproal index and $\beta$ is spectral index.
 
  Before determining the SED at this epoch,  we consider if it is appropriate to consider such a SED 
 as a simple power law SED from synchrotron radiation of afterglow by analyzing their temporal evolutions.
  At $t > 1500$ sec, P10 find that the light curves in the X-ray, and the optical/NIR are
  described by simple power-laws. By fitting single power-law functions to the light curves of the
  publicly available $Swift$ XRT data \footnote{http:$//$www.swift.ac.uk$/$xrt$\_$curves$/$00295301$/$}\citep{eva07} and to the $I$, $J$, and $K_{s}$-band data including those presented in
  P10, we find that over an interval of $t =$ 1900 to 17000 sec $\alpha_{X} = 1.75 \pm 0.06$, $\alpha_{I}$ = $1.50 \pm 0.05$, $\alpha_{J}$ = $1.44 \pm 0.04$, and $\alpha_{K}$ = $1.41 \pm 0.03$,
  or $\alpha_{optical/NIR} - \alpha_{X} \simeq -0.25$ to  $-0.34 \pm 0.08$.
   When the analysis is extended to $t \sim 10^{5}$ sec, we get $\alpha_{J} - \alpha_{X} = -0.27 \pm 0.04$.
  These results, being consistent with $\alpha_{optical} - \alpha_{X} < -0.25$ of uniform ambient environment \citep{ura07}, suggest that cooling break ($\nu_c$) lies between the X-ray and the optical wavelengths and that the light in optical/NIR is dominated by
 the afterglow from external shock.  Since we also find no late time prompt signal
 (e.g., flare and shallow decay) in X-ray, we conclude that stable afterglow light characterizes 
 the SED at $t = 6605$ sec.

The remaining task is to synchronize $R$ and $I$ band photometry to this epoch. One of the $I$ band images from LOAO is nearly contemporaneous to the $JHK_{s}$ data, lagging behind the NIR observation only by 209 sec. To correct for the small difference in the observed time, we interpolate the $I$ band data with the power law model.
 
We fit the $I$ band light curves only with our own data and with additional $I$ band data from P10. The fitted values of $\alpha$ are $1.48 \pm 0.11$ and $1.48 \pm 0.04$  without or with the P10 data respectively. The necessary correction to derive the contemporaneous $I$ band magnitude from the 6814 sec post-burst data point is +0.050 mag with $\alpha=1.48$. Because the choices of $\alpha$ in both two cases share the same value in spite of different errors, we will adopt the correction of 0.05 mag to compute the $I$ band magnitude at 6605 sec, which gives $I = 19.18 \pm 0.16$ mag. For $R$ band photometry, we used two methods. First, we extrapolated the $R$ band detection at $t$ = 1500 sec to $t$ = 6605 sec assuming an achromatic decay of the light curve during this period as found in P10. This gives the $R$ band photometry of $R$ = 21.3 mag. As the second approach, we derived a limit in the $R$ band photometry from a stacked image made of seven 300 sec $R$ band frames around $t$ = 6605 sec. In the stacked image, no afterglow was detected, giving an upper limit in $R$ band photometry at 21.1 mag at 3 $\sigma$. These two methods give consistent results, therefore we adopt $R$ = 21.3 mag for the $R$ band photometry at $t$ = 6605 sec. For the 1 $\sigma$ uncertainties of data, 0.17 mag is used in $I$ band to reflect the uncertainty in $\alpha$. In the case of $R$ band, we adopt the error, 0.3 mag, of the $R$ frame at $t$ = 1500 sec.\\

\section{Analysis Method} \label{bozomath}
 The observed photometry data are fitted using model SEDs as described below. The underlying afterglow emission is assumed to follow a simple power law of $F_{\nu} \sim \nu^{-\beta}$ as produced by synchrotron emission of relativistically  moving particles. The attenuation of photons below Ly$-\alpha$ is implemented following the prescription of  \citet{mad95}. For the intrinsic dust extinction of GRB 071025, we try the Milky Way (hereafter, MW), LMC, SMC extinction laws \citep{pei92}, and an extinction curve from \citet{cal94} which is usually applied to star-forming galaxies. For SNe-dust extinction curves, we tried an empirical extinction curve derived from $z \sim 6$ quasars \citep{mai04}, and theoretical extinction curves of H05 and H08 (see discussion in Section 4.2 for more detail of the model). The model SEDs were convoluted with the transmission curves of filters and quantum efficiency of CCD \citep{lee10}.

\section{Resutls \& Discussion}
\subsection{Redshift of GRB 071025} \label{bozomath}
We derive the redshift of GRB 071025 before proceeding to a detailed comparison of the best-fit models using different extinction curves. By fixing the redshift, the direct comparison between different models becomes more straightforward. Fixing the redshift to a single value can be justified if the derived redshift is not strongly dependent on models.

In order to determine the redshift, we fit the observed SED with different extinction curves with four free parameters: redshift, $A_{V}$, beta, and the normalization factor. We find that the derived redshifts are all 
betwen 4.6 to 4.85
due to the sharp break at $R$ band with $R - I \simeq 2.0$ which can be naturally explained with intergalactic absorption of the light below Ly-$\alpha$. The result is consistent with the value reported in P10.

 If we change the redshift 2 $\sigma$ apart from the best fit result for each extinction curve, the amounts of increased $\chi^{2}/\nu$ values for SN-dust extinction curves approximately range from 3 to 5 : 3.48 (Maiolino), 4.21 (Hirashita 25M$_{\odot}$), and 5.46 (Hirashita 13M$_{\odot}$) whereas other extinctions show higher values than SN-dust extinctions do : 7.90 (SMC), 9.12 (LMC), and 12.40 (Calzetti). The MW extinction  shows a larger increase, 16.23 compared to the others, which can be explained by the influence from 2175\AA$\;$bump. Within 2 $\sigma$ deviation of redshit from the best fit, SN-dust curves always present less $\chi^{2}/\nu$ than the other extinctions. Consequently, none of all extinctions have better fitting results when the redshift deviated by 2 $\sigma$ from the best fitting one is chosen.
Since the dependence of the redshift on the assumed dust extinction curves is small, we will fix the redshift of GRB 071025 to be a fiducial value of 4.8 in the following analysis of the extinction property. We also tried the same analysis using different values for the redshift, and found that the conclusion reached in the paper is not affected by the choice of the redshift.
 The difference in the reduced $\chi^{2}$ values between different models does not 

%

\subsection{Dust Properties} \label{bozomath}

 Figure 1 shows the result of the SED fit with various dust extinction curves (SMC, MW, Calzetti, and Maiolino). Like in P10, we observe an inflextion in the observed SED shape -- As we go from the shorter wavelength to the longer wavelength, $IJHK_{s}$ data points move up and down with respect to a simple power law model SED. While the best-fit models with SNe-dust have the $\chi^{2}/\nu$ values less than 1, models with the dust extinction curves of LMC, SMC, MW, and the Calzetti produce the $\chi^{2}/\nu$ values of about 4 $-$ 5. The MW and the LMC extinctions are characterized by a 2175\AA$\;$bump which is absent in the SED of GRB and the best-fit $A_{V}$'s of these models converge to zero. The SED with the Calzetti curve is featureless, and similar to the SED with no extinction. Although the adoption of the SMC extinction curve improves the goodness of the fit a little bit compared to the MW or the Calzetti curves, it cannot reproduce the observed changes of slopes between $JHK_{s}$ bands as well as the SNe-dust extinction curves.

Since some of the fits converge to no extinction, we tried a fit without dust extinction, which has a benefit of reducing one free parameter. The fit returns the $\chi^{2}/\nu$ of 2.55, an improvement over the fits with the MW and the LMC extinction curves. However, the best-fit spectral slope is too steep at $\beta = 1.6$ 
 in comparsion with $\beta_{X}$ $\sim$ 1 observed by the XRT, which is inconsistent with the synchrotron theory, $\beta_{optical,IR} = \beta_{X} - 0.5$ if $\nu_{c}$ is between X-ray and optical/IR. 
 As our light curve analysis suggests that the afterglow emission follows the external shock model with 
 $F(t,\nu) \sim t^{\frac{3(p-1)}{4}}\, \nu^{\frac{p-1}{2}}$ for the optical/NIR where $p$ is electron power
 law index \citep{sar98}. From our optical/NIR light curves, we get  
 $p \sim 2.87$ and consequently $\beta \sim 0.94$. Also, GRB afterglows are known to
 have $\beta \sim 0.6$ and $\beta \lesssim 1.1$ for all the well-studied cases \citep{kan06, kan10, zar11}.
 Therefore, we argue that no extinction model is unfavorable.

 The SNe-dust models perform far better than the models with the other dust extinction laws,
 confirming the previous result from P10.

\subsection{Progenitor and Environment of the Dust-Producing SNe} \label{bozomath}
 The H05 and H08 allow us to test various SNe-dust production scenarios with a broad progentior mass range from 13 to 170M$_{\odot}$, with or without mixing of heavy elements inside the He-shell, and with or without dust destruction due to the reverse shock. Dust grains can be destroyed by the SNe shock, modifying the grain size distribution and eventually the extinction curve. In H08, the shock destruction of the dust within SNe is implemented, and the destruction efficiency depends on the ambient gas density and it is varied from 0, 0.1, and 1 cm$^{-3}$, with 0 being the case for no dust destruction and 1 cm$^{-3}$ for the most efficient dust destruction. If the gas density is much larger than 1 cm$^{-3}$, the dust destruction is so efficient that there would be almost no dust extinction \citep{noz07}. Figure 2 and Table 2 summarize our finding after the GRB 071025 data fitted with these models.

 First, we discuss which of the mixed or the unmixed core model fits the data better using 25M$_{\odot}$ progenitor models as references. The right panel of Figure 2 shows that the unmixed core model fits the data better than the mixed core model. Mixed core only produces oxidized molecules, since most carbons are locked in CO. Extinction is dominated by SiO$_{2}$ in such cases, regardless of progenitor mass. Since the SiO$_{2}$ -- dominated extinction curve monotonically increases with 1/$\lambda$ without any change of the slope, it cannot explain the inflexing SED shape of GRB 071025. We find that the same argument applies to models with different progenitor masses, therefore, our discussion to unmixed core progenitors in the following.

 Next, we examine if the dust destruction is needed to account for the observed SED. The triple dot-dashed and the dotted lines represent models with dust destruction at n$_{H}$=0.1 cm$^{-3}$ and 1 cm$^{-3}$ respectively. 
The destruction affects small grains, and the effect is reflected in the model SED as shown in the left panel of Figure 2. After the destruction, the extinction curve becomes featureless resembling the SMC or the Calzetti curves, and the afterglow model SEDs with the dust destruction effect fit the data as poor as those with the SMC or the Calzetti curve. Therefore, we conclude that dust destruction should be negligible for GRB 071025. We checked that this conclusion holds for models with other progenitor masses.

 Finally, we present what the most likely progenitor mass is for the SNe responsible for the dust. For this, we compare the best-fit models 
 with 13, 20, 25, 30, and 170M$_{\odot}$ progenitors with unmixed core and no dust destruction (right panel of Figure 2). 
 Figure 2 shows that the models with the progenitor mass of 13 or 25M$_{\odot}$ fit the data significantly better than   
 the others (the $\chi^{2}/\nu \lesssim 1$ versus $\sim 5$). This is because the inflexing SED shape between $I$ and $K_{s}$ bands can be explained by extra absortions at the rest-frame 1200 $-$ 2000\AA $\ $ and at around 3000\AA$\ $, which results from the increased contribution of carbon grains (13M$_{\odot}$ model), or Mg$_{2}$SiO$_{4}$ and FeS grains (25M$_{\odot}$ model) with respect to Si grains. The 30M$_{\odot}$ progenitor model fails to fit the data because of the excessive increase of Si grains. The 170M$_{\odot}$ model, given as a representative case for Pair Instability Supernovae (PISNe) which could be abundant in the early universe \citep{heg02}, also gives poor fits to our data. This results from the high contribution of Si grains that make the curve look more or less monotonically increasing, and the deficiency of the extra extinction from Fe that introduces a plateau in the model SED at shorter wavelengths. 

Additionally, in order to confine the most optimal progenitor mass of SN-dust extinction for GRB 071025, we introduce the $Y$ band data from P10. We interpolated the $Y$ data from REM and MAGNUM in P10 to determine $Y$ magnitude at $t$ = 6605 sec through \citet{hil02}. Although the $Y$ magnitude was inserted into the SED, both Hirashita 13 and 25M$_{\odot}$ models show acceptable fitting, as $\chi^{2}/\nu$ = 2.48/(5-3) and 0.57/(5-3) for 13 and 25M$_{\odot}$, respectively. When the dust destruction is taken into consideration, the $Y$ magnitude do not still enhance the poor fits of models where dust destructions exist. In conclusion, the additional information of Y do not alter the consequences of our own dataset for appropriate SN-dust extinctions.

\section{Summary}
We examined the dust property of a red afterglow of GRB 071025. Analysis of our own dataset which is largely independent from P10 supports the evidence for the existence of SNe-dust at z $\sim$ 5. Furthermore, using the SNe-dust extinction curves of H05 and H08, we find that models with 13 and 25M$_{\odot}$ progenitors with unmixed core and no dust destruction give the best fit to the data. Given uncertainties in the model, we suggest that the most plausible progenitors of the SNe-dust are intermediate mass stars with 10 $-$ 30M$_{\odot}$. 

\acknowledgments
We thank R. Chary and D. Perley for useful discussions. This work was supported by the Korea Science and Engineering Foundation (KOSEF) grant No. 2009-0063616, funded by the Korea government (MEST). We acknowledge the use of data obtained with LOAO 1-m telescope operated by KASI and the Canada-France-Hawaii Telescope, and data supplied by the UK Swift Science Data Centre at the University of Leicester. This work is partly supported by NSC-99-2112-M-008-003-MY3(YU) and NSC-99-2112-M-001-002-MY3(KYH). Access to the CFHT was made possible by the ASIAA, Taiwan.

\clearpage

\begin{figure}
\centering
\includegraphics[angle=0,width=1.\textwidth]{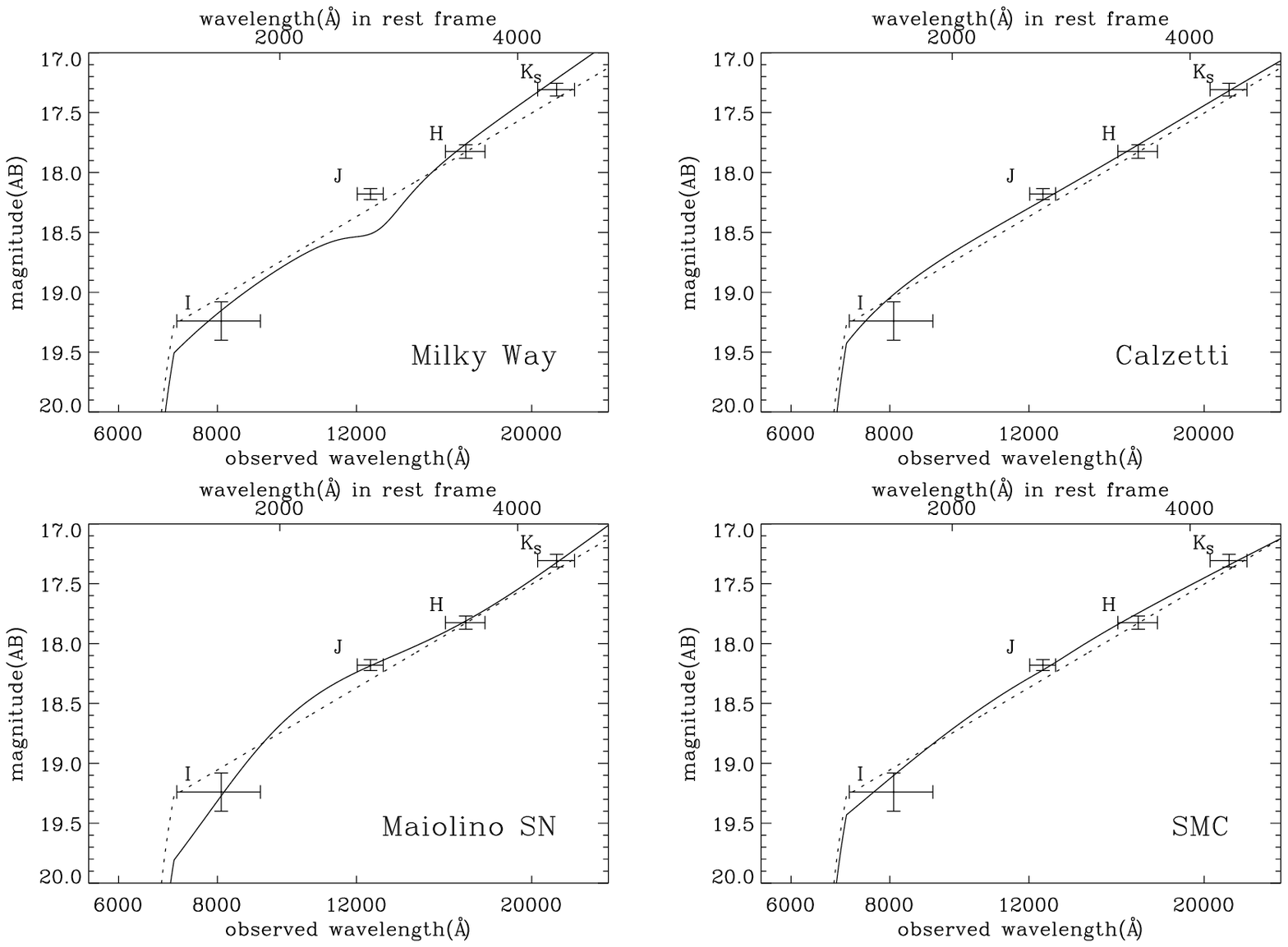}
\caption{SEDs of GRB 071025 at the 6605 sec post-burst epoch, fitted with models incorporating various extinction curves. The solid lines indicate the best-fit model curves, while the dashed line shows a result with no extinction. The Maiolino model, given as a representative of the SNe-dust extinction curves, fits the data best. The $dotted$ lines for each panel are plotted with no extinction assumption.\label{figure1}}
\end{figure}

\clearpage

\begin{figure}
\centering
\includegraphics[angle=0,width=1.\textwidth]{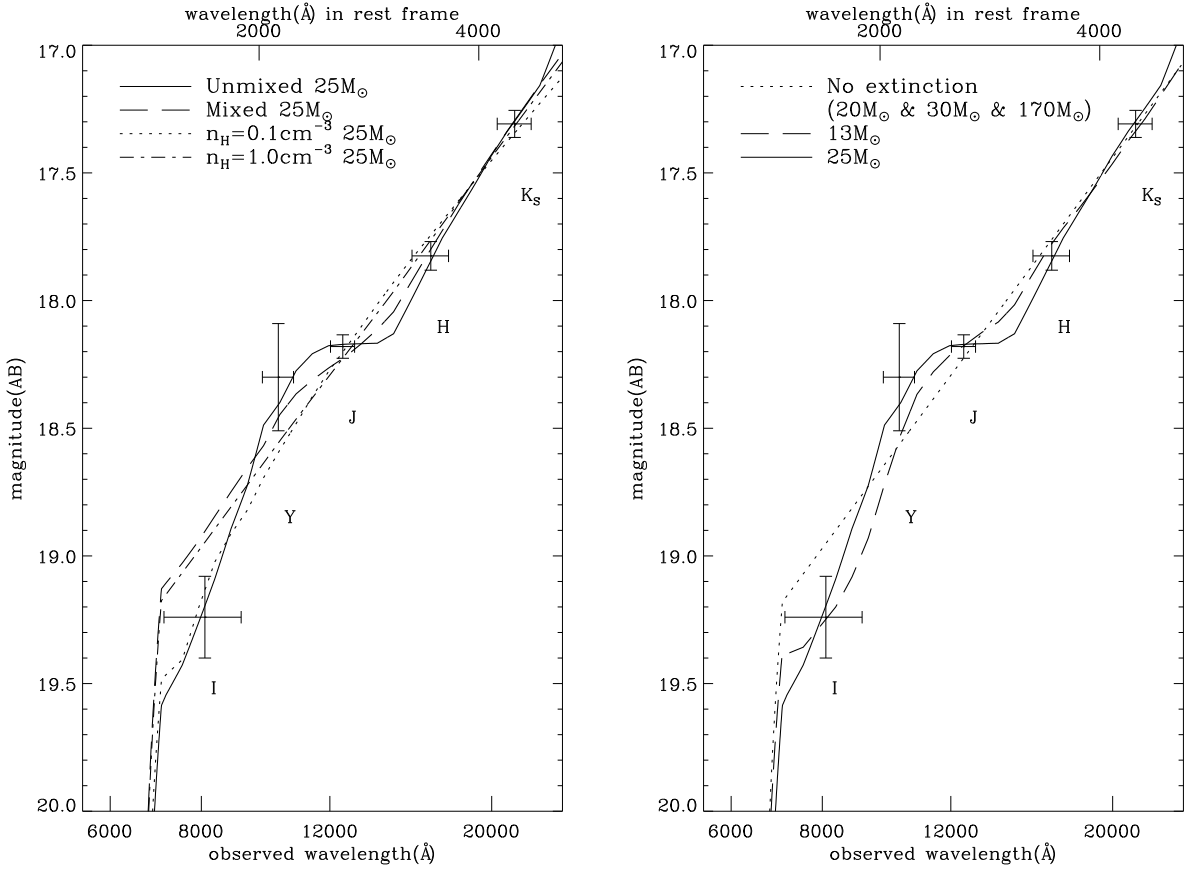}
\caption{Comparison of the GRB 071025 SED with SNe-dust models with different physical ingredients such as dust destruction, progenitor mass, and mixed/unmixed core. In the left panel, we show 25M$_{\odot}$ progenitor models with mixed core (the long-$dashed$ line), with unmixed core (the $solid$ line), and with unmixed core with dust destructions where the ambient densities are n$_{H}$=0.1 cm$^{-3}$ and n$_{H}$=1.0 cm$^{-3}$. In the right panel, we indicate cases of different progenitor masses with unmixed cores under no dust destruction. In both figures, the $dotted$ lines indicate cases without dust extinction which are virtually identical to cases with very efficient dust destructions. The figure shows that the models with unmixed cores, 13 and 25M$_{\odot}$ progenitors, and no dust destruction fit the data best. $Y$ magnitude is extracted from P10. \label{figure2}}
\end{figure}

\clearpage

\begin{deluxetable}{ccccr}
\tablewidth{0pt}
\tablecaption{Optical \& NIR Photometry of GRB 071025}
\tablehead{\colhead{Filter} & \colhead{Facility} & \colhead{Mid-point time elapsed} & \colhead{Total integration time} & \colhead{Magnitude $^{a,b}$}\\
\colhead{} & \colhead{} & \colhead{after $Swift$ alert} & \colhead{} & \colhead{}}
\startdata
I & LOAO &1091 s & 30 s & 17.33 $\pm$ 0.08 \\
I & LOAO &1901 s & 300 s & 17.27 $\pm$ 0.09 \\
I & LOAO &3703 s & 300 s & 18.46 $\pm$ 0.09 \\
I & LOAO &5061 s & 300 s & 18.81 $\pm$ 0.15 \\
I & LOAO        & 6605 s & .       &  19.18 $\pm$ 0.16\\
I & LOAO &6814 s & 300 s & 19.23 $\pm$ 0.16 \\
R & LOAO &1563 s & 300 s & 18.95 $\pm$ 0.17 \\
R & LOAO &6605 s & . & 21.3 $\pm$ 0.30\\
J &  SQIID &6605 s & 240 s & 18.18 $\pm$ 0.05 \\
H & SQIID &6605 s & 240 s & 17.83 $\pm$ 0.06 \\
K & SQIID &6605 s & 240 s & 17.31 $\pm$ 0.05 \\
J & CFHT & 90589 s & 1200 s & 22.13 $\pm$ 0.35 \\
H & CFHT & 92308 s & 1050 s & 21.56 $\pm$ 0.24 \\
K & CFHT & 88792 s & 1200 s & 20.78 $\pm$ 0.12 \\

\enddata
\tablenotetext{a}{The magnitudes are in AB magnitude system}
\tablenotetext{b}{Galactic extinction was corrected}
\end{deluxetable}

\clearpage

\begin{deluxetable}{lccc}
\tablewidth{0pt}
\tablecaption{Fitting Results Varying Extinction Curves}
\tablehead{\colhead{Extinction curves $^{a}$} & \colhead{$\beta$} & \colhead{$A_{3000}$} & \colhead{$\chi^{2}$/$\nu$}}
\startdata
None & 1.56$\pm$0.11 & . & 5.10/(4-2)\\
Milky Way & 1.56$\pm$0.11 & $\thicksim$0 & 5.10/(4-3)\\
LMC & 1.56$\pm$0.11 & $\thicksim$0 & 5.10/(4-3)\\
SMC & 1.05$\pm$0.47 & 0.30$\pm$0.24 & 3.87/(4-3)\\
Calzetti & $\thicksim$0 & 2.57$\pm$0.07 & 3.94/(4-3)\\
Maiolino SNe & 1.24$\pm$0.17 & 0.77$\pm$0.34 & 0.17/(4-3)\\
Hirashita 13M$_{\sun}$(U) & 0.86$\pm$0.32 & 1.73$\pm$0.77 & 2.48/(5-3)\\
Hirashita 20 \& 30M$_{\sun}$(U) & 1.54$\pm$0.10 & $\thicksim$0 & 7.29/(5-3)\\
Hirashita 20M$_{\sun}$(M) & 1.19$\pm$0.37 & 0.15$\pm$0.15 & 6.44/(5-3)\\
Hirashita 25M$_{\sun}$(U) & 0.79$\pm$0.30 & 2.65$\pm$1.00 & 0.57/(5-3)\\
Hirashita 25M$_{\sun}$(M) & 1.26$\pm$0.23 & 0.07$\pm$0.05 & 5.40/(5-3)\\
Hirashita 25M$_{\sun}$ $n_{H}$=0.1(U) & 1.36$\pm$0.20 & 0.77$\pm$0.73 & 6.37/(5-3)\\
Hirashita 25M$_{\sun}$ $n_{H}$=1.0(U) & 1.53$\pm$0.16 & 0.09$\pm$0.88 & 7.26/(5-3)\\
Hirashita 170M$_{\sun}$(U \& M) & 1.54$\pm$0.10 & $\thicksim$0 & 7.29/(5-3)\\
Hirashita 170M$_{\sun}$ $n_{H}$=0.1(U) & 1.54$\pm$0.10 & $\thicksim$0 & 7.29/(5-3)\\
Hirashita 170M$_{\sun}$ $n_{H}$=1.0(U) & 1.54$\pm$0.10 & $\thicksim$0 & 7.29/(5-3)\\
\enddata
\tablenotetext{a}{U:Unmixed core, M:Mixed core}
\end{deluxetable}

\end{document}